\documentclass[11pt,letterpaper]{article}
\usepackage{jcappub}
\usepackage{bbm}
\usepackage{mathrsfs}
\usepackage{slashed}
\usepackage{caption}
\usepackage{epstopdf}
\usepackage[normalem]{ulem}
\usepackage[bottom]{footmisc}
\usepackage{subcaption}
\usepackage{bbold}
\usepackage{titlesec}
\usepackage{threeparttable}
\usepackage{booktabs}
\usepackage{changepage}
\usepackage[utf8]{inputenc}
\usepackage{dsfont} 
\usepackage{grffile}
\usepackage{graphicx}  % needed for figures
\usepackage{dcolumn}   % needed for some tables
\usepackage{bm}        % for math
\usepackage{amssymb}   % for math
\usepackage{setspace}
\usepackage{amsmath, amssymb, setspace}
\usepackage{array}
\usepackage{booktabs}
\usepackage{caption}
\usepackage{indentfirst}
\usepackage{float}
\usepackage{lmodern}
\usepackage{multirow}
\usepackage{soul}
\usepackage[normalem]{ulem}
\usepackage{braket}
\usepackage{comment}
\usepackage[draft]{pgf}
\usepackage{adjustbox} 
\newcommand{\DoBox}[1]{\begin{center}
\color{red}\fbox{
\begin{minipage}{0.9\textwidth}
\end{minipage}}
\end{center}}

\newlength{\myimageoversize}
\newsavebox{\myimage}

\usepackage[most]{tcolorbox}
\usepackage{empheq}

\tcbset{colback=white!10!white, colframe=black!50!black, 
        highlight math style= {enhanced, %<-- needed for the ’remember’ options
            colframe=black,colback=white!10!white,boxsep=0pt}
        }

\begin{document}
\title{\huge{Testing the ALP-photon coupling with polarization measurements of Sagittarius A$^\star$}}

\author{Guan-Wen Yuan$^{a,b}$,}
\author{Zi-Qing Xia$^{a}$\footnote{Corresponding author.},}
\author{Chengfeng Tang$^{b,c,d}$,}
\author{Yaqi Zhao$^{b,c,d}$,}
\author{Yi-Fu Cai$^{b,c,d}$,}
\author{Yifan Chen$^{e}$\footnote{Corresponding author.},}
\author{Jing Shu$^{e,f,g,h,i,j}$,}
\author{Qiang Yuan$^{a,b,h}$,}
%\author{Yue Zhao$^k$}

\emailAdd{yuangw@pmo.ac.cn}
\emailAdd{xiazq@pmo.ac.cn}
\emailAdd{tcf123@mail.ustc.edu.cn}
\emailAdd{zxmyg86400@mail.ustc.edu.cn}
\emailAdd{yifucai@ustc.edu.cn}
\emailAdd{yifan.chen@itp.ac.cn}
\emailAdd{jshu@itp.ac.cn}
\emailAdd{yuanq@pmo.ac.cn}
%\emailAdd{zhaoyue@physics.utah.edu}

\affiliation{
$^a$Key Laboratory of dark Matter and Space Astronomy, Purple Mountain
Observatory, Chinese Academy of Sciences, Nanjing 210033, China \\
$^b$School of Astronomy and Space Science, University of Science and Technology of China, Hefei, Anhui 230026, China\\
$^c$Department of Astronomy, School of Physical Sciences, University of Science and Technology of China, Hefei, Anhui 230026, China\\
$^d$CAS Key Laboratory for Researches in Galaxies and Cosmology, University of Science and Technology of China, Hefei, Anhui 230026, China\\
$^e$CAS Key Laboratory of Theoretical Physics, Institute of Theoretical Physics, Chinese Academy of Sciences, Beijing 100190, China\\
$^f$School of Physical Sciences, University of Chinese Academy of Sciences, Beijing 100049, China\\
$^g$CAS Center for Excellence in Particle Physics, Beijing 100049, China\\
$^h$Center for High Energy Physics, Peking University, Beijing 100871, China\\
$^i$School of Fundamental Physics and Mathematical Sciences, Hangzhou Institute for Advanced Study,
University of Chinese Academy of Sciences, Hangzhou 310024, China\\
$^j$International Center for Theoretical Physics Asia-Pacific, Beijing/Hanzhou, China\\
%$^k$Department of Physics and Astronomy, University of Utah, Salt Lake City, UT 84112, USA
}
\date{\today}

\abstract{
Ultra-light bosons such as axions or axion-like particles (ALPs), are promising candidates to solve the dark matter problem. A unique way to detect such ALPs is to search for the periodic oscillation feature of the position angles of linearly polarized photons emitted from the galaxy center. In this work, we use the high-resolution polarimetric  measurements of the radiation near the super-massive black hole (SMBH) in the center of the Milky Way, i.e., Sagittarius A$^\star$ (Sgr A$^\star$), by a sub-array of the Event Horizon Telescope to search for the ultra-light ALPs. We derive upper limits on the ALP-photon coupling of $\sim 10^{-12}~{\rm GeV^{-1}}$ for ALP masses of $m_a\sim (10^{-19}-10^{-18})$~eV, with a solitonic core + NFW dark matter density profile. Our results are stronger than that derived from the observations of SN1987A and a population of supernovae in the mass window of ($10^{-19}-10^{-17}$)~eV. Improved polarimetric measurements with the full Event Horizon Telescope can further strengthen the constraints. 
}
% \vspace{5mm}
% Keywords: axion, black hole, soliton core
\maketitle

\section{Introduction}
As a crucial component of the Universe, the nature of dark matter (DM) remains a mystery \cite{Rubin:1980zd, Clowe:2006eq, Larson:2010gs}. For quite a long time the weakly interacting massive particles (WIMPs) are regarded as the most prevailing hypothetical candidates of DM due to the success in explaining the relic abundance of DM \cite{Bertone:2004pz}. However, so far there is no convincing evidence of the existence of WIMPs in many direct detection experiments and very stringent limits on the cross section between DM particles and standard model particles have been given \cite{Feng:2010gw,Liu:2017drf,Zhao:2020ezy}. The experimental searches for other candidates of DM have thus attracted more and more attention in recent years \cite{Sigl:2018fba,Bertone:2019irm}. As one of the many alternatives, a class of ultra-light bosonic particles, axions and their extended scenarios dubbed as axion-like particles (ALPs), have been widely discussed due to their broad connections with several important problems in physics and cosmology. This type of hypothetical pseudo-scalar fields were originally motivated by the solution of the strong-CP problem in quantum chromodynamics (QCD) \cite{Peccei:1977hh, Weinberg:1977ma, Wilczek:1977pj, Dine:1981rt}. While the favored mass range of the QCD axions may be limited, and the mass-coupling relation is largely fixed, the ALP scenario can span a much wider parameter range of the mass and coupling, and can be viable DM candidates \cite{Marsh:2015xka}. Very interestingly, ultra-light ALPs (with $m_a\sim10^{-22}$ eV), which are naturally predicted in the string theory \cite{Arvanitaki:2009fg}, can potentially solve the ``small-scale crisis'' of the structures in the cold DM Universe \cite{Preskill:1982cy, Abbott:1982af, Dine:1982ah, Hu:2000ke}, and are regarded as one of the most promising DM candidates other than WIMPs.

There have been many experimental designs searching for axions or ALPs based on their possible coupling to the electromagnetic sector. These include, for example, the conversion of ALPs into photons \cite{Anastassopoulos:2017ftl, Payez:2014xsa,Meyer:2020vzy,Chris:200803305}, the spectral distortion of photons \cite{Hooper:2007bq, TheFermi-LAT:2016zue, xia2018prd,liang2019JCAP, Xia:2019yud, Li:2020pcn}, the fifth force mediated by ALPs \cite{Jaeckel:2010ni, Rong:2017wzk, Rong:2018yos} 
and cosmological bound \cite{Dror:2020zru}. All these experiments have already excluded a sizable parameter space of the ALPs \cite{Graham:2015ouw}. 

Another independent and complementary method is that the axion field can give rise to a rotation of the photon polarization position angle due to its interaction with electromagnetic field, which is similar with the classical Faraday rotation effect and known as one kind of birefringence effect \cite{Carroll:1989vb, Carroll:1991zs, Harari:1992ea, Ivanov:2018byi, Fujita:2018zaj, Liu:2019brz, Fedderke:2019ajk, Caputo:2019tms, Chen:2019fsq, Chigusa:2019rra, Poddar:2020qft, Basu:2020gsy}. Different from the axion-photon conversion, the birefringence effect is path-independent and frequency-independent. The variation amplitude of the position angle is proportional to the axion field strength, and thus radiation from where high density axion fields form, for example, those located in the galactic halo center or around massive black holes, are ideal targets to search for this birefringence effect. High-resolution observations, e.g., with the fabulous technology of the very-long-baseline interferometry (VLBI) \cite{Akiyama:2019brx, Middelberg:2008qc, 2018A&A616L15X}, are usually required, since we want to observe small regions in the center of a galaxy. Recently, the successful operation of the Event Horizon Telescope (EHT) program, which is a global VLBI array operating at mm wavelengths, has made it possible to observe the detailed polarimetric images of the nearby super-massive black holes Messier 87 (M87$^\star$) and Sagittarius A$^\star$ (Sgr A$^\star$) \cite{Doeleman:2008qh, Johnson:2015iwg, Johannsen:2016vqy}. These high-resolution polarimetric observations provide us with a powerful and unique way to search for ALPs.

In this work, we search for the ALP-photon coupling signature using the position angle variations of polarized millimeter emission from Sgr A$^\star$ observed by a sub-array of EHT \cite{Johnson:2015iwg}. For the density profile of ALPs in the Galactic center, we adopt a solitonic core at small radii plus a Navarro-Frenk-White (NFW) profile at large radii \cite{Navarro:1996gj}. We use the maximum likelihood method to search for a periodic oscillation of the position angle in the data. Since the measurement results at different time show relatively large variations which may be due to the astrophysical environmental effects, we choose the data taken in Day 82 of Ref. \cite{Johnson:2015iwg} which are relatively stable within a period of $\sim 4$ hours to give the constraints on the ALP-photon coupling. The observation time coverage of $\sim 4$ hours with a cadence of $\sim20$ minutes is sensitive to ALPs with mass of $\sim10^{-18}$~eV.

The rest of this paper is organized as follows. In Section \ref{theory part}, we review the photon propagation in an external ALP background, and derive the corresponding position angle variation of a linearly polarized photon due to the coupling with the oscillating ALP field. In Section \ref{data analysis}, we present the constraints on the ALP parameters from the likelihood analysis of the polarimetric measurement data of Sgr A$^\star$. We give some discussion of the current work and future prospects in Section \ref{discussion}, and summarize this work in Section \ref{summary}.

\section{Birefringence from ALPs}\label{theory part}

\subsection{ALP-photon interaction}
The ALP field can interact with the electromagnetic field, giving rise to periodic rotation of the position angle of a linearly polarized photon, which is known as the birefringence effect. The relevant Lagrangian terms include
\begin{equation}
\label{Lagrangian}
    \mathcal{L}=-\frac{1}{4} F_{\mu \nu}F^{\mu \nu}+\frac{1}{2}\left(\partial_{\mu} a \partial^{\mu} a-m_a^{2} a^{2}\right)+\frac{g_{a \gamma}}{4} a F_{\mu \nu} \tilde{F}^{\mu \nu}
\end{equation}
where $a$ is the ALP field with mass $m_a$, $F_{\mu\nu}$ is the electromagnetic stress tensor and $g_{a\gamma}$ is the coupling constant between the ALP and the photon field. The equations of motion arising from Eq.~(\ref{Lagrangian}) are
\begin{equation}\label{photon_EOM}
\ddot{\boldsymbol{A}}-\nabla^{2}\boldsymbol{A}=g_{a \gamma}(\dot{a}\nabla\times\boldsymbol{A}+\dot{\boldsymbol{A}} \times\nabla a)
\end{equation}
and
\begin{equation}\label{axion_EoM}
\square a+m_a^{2} a=g_{a \gamma} \boldsymbol{E} \cdot \boldsymbol{B}.
\end{equation}
Assuming $\left|g_{a \gamma} \boldsymbol{E} \cdot \boldsymbol{B}\right| \ll m_a^{2}a$, we can ignore the backaction term. Then a simple solution to Eq.~(\ref{axion_EoM}) is the coherently oscillating ALP field
\begin{equation}
a(t,\boldsymbol{x})=a_0(\boldsymbol{x}) \sin[m_at+\delta(\boldsymbol{x})],
\end{equation}
where $a_0(\boldsymbol{x})$ is the amplitude and $\delta(\boldsymbol{x})$ is the position-dependent phase. We assume that the time variation scale of $a_0(\boldsymbol{x})$ and $\delta(\boldsymbol{x})$ are much longer than the oscillation period of the ALP field
\begin{equation}
\label{oscillation_period}
    T=\frac{2\pi}{m_a}\simeq 4\times 10^{3}\left(\frac{10^{-18}~{\rm eV}}{m_a}\right)~\textrm{sec}.
\end{equation}
so that we can treat them as time-independent. Following \cite{Fujita:2018zaj}, in the temporal gauge and the Coulomb gauge, $\boldsymbol{A}(t, \boldsymbol{x})$ can be decomposed into two circular polarization modes $A^{\pm}$:
\begin{equation}
  \boldsymbol{A}(t, \boldsymbol{x})=\sum_{\pm} \int \frac{d^{3} k}{(2 \pi)^{3}} A^{\pm}(k) \boldsymbol{e}^{\pm}(\hat{\boldsymbol{k}})e^{( ik \cdot x- i\omega_{\pm}t)},
\end{equation}
with unit circular polarization vectors $\boldsymbol{e}^{\pm}(\hat{\boldsymbol{k}})$ satisfying 
$i \boldsymbol{k} \times \boldsymbol{e}^{\pm}(\hat{\boldsymbol{k}})=\pm k \boldsymbol{e}^{\pm}(\hat{\boldsymbol{k}})$. Taking the decomposition into the equation of motion (\ref{photon_EOM}), we can get the two different dispersion relations for the corresponding circular polarized modes
\begin{equation}
\label{dispersion_relation}
    w^2_{\pm}-k^2 \mp g_{a \gamma}(\dot{a}+\hat{\boldsymbol{k}} \cdot \boldsymbol{\nabla} a)|k|=0.
\end{equation}
In the limit $k \gg m_a$, the solutions reduce to $w_{\pm}\simeq k \pm\frac{g_{a\gamma}}{2}\frac{da}{dt}$ with $d/dt$ being the total derivative along the propagation path of the photon. This birefringence comes from the spontaneous breaking of the parity symmetry due to the misalignment of the ALP field. For a linearly polarized photon, this leads to the shift of the position angle $\phi$ in the polarization plane with
\begin{equation}
\begin{aligned}
\label{delta_angle1}
    \Delta \phi&=\frac{1}{2}\int_{t_{\rm emit}}^{t_{\rm obs}} (w_{+}-w_{-} )d t= \frac{g_{a \gamma}}{2} \int_{t_{\rm emit}}^{t_{\rm obs}} \frac{da}{dt}  d t\\ &=\frac{g_{a\gamma}}{2}[a(t_{\rm obs},\boldsymbol{x}_{\rm obs})-a(t_{\rm emit},\boldsymbol{x}_{\rm emit})],
\end{aligned}
\end{equation}
where $a (t_{\rm emit},\boldsymbol{x}_{\rm emit})$ and $a (t_{\rm obs},\boldsymbol{x}_{\rm obs})$ are the ALP field at the emission and the observation point respectively. We will be interested in the situation where the density of the ALP field in the vicinity of the light source is much higher than the one near the observer, i.e. $a_0 (t_{\rm emit},\boldsymbol{x}_{\rm emit})\gg a_0 (t_{\rm obs},\boldsymbol{x}_{\rm obs})$. Thus only the ALP field around the emission point becomes relevant. Assuming that ALPs make up the dominant DM and using the relation $\rho_{\mathrm{DM}}=\frac{1}{2} m_a^{2} a_{0}^{2}$, the position angle shift can be parameterized as 
\begin{equation}
\label{delta_angle2}
\begin{aligned}
    \Delta \phi (t) \simeq 5^{\circ} \sin \left(2 \pi \frac{t}{T}+\delta(\boldsymbol{x})\right)\left(\frac{\rho_{\mathrm{DM}}}{2\times 10^{9}~\mathrm{GeV} / \mathrm{cm}^{3}}\right)^{\frac{1}{2}}\left(\frac{g_{a \gamma}}{10^{-12}~ \mathrm{GeV}^{-1}}\right)\left(\frac{m_a}{10^{-18}~\mathrm{eV}}\right)^{-1}.
\end{aligned}
\end{equation}
The final position angle contain two parts
\begin{equation}
\label{theorial_phi}
    \phi (t) = \Delta\phi (t) + \phi_{\rm bkg},
\end{equation}
where $\Delta\phi (t)$ is the ALP induced position angle shift (signal), which can be seen as a slowly varying perturbation for the observation, and $\phi_{\rm bkg}$ is the position angle shift due to the astrophysical background (e.g., the varying magnetic field). 

\subsection{DM density distribution}
At regions far away from the galaxy center, the DM density distribution can be approximately described by an NFW profile. For ultralight axion DM, a solitonic core can form in the galaxy center, due to the balance between the gravitational interaction and the quantum pressure. The radius of the solitonic core $r_c$ is related to the de Broglie wavelength of the ALP field
\begin{equation}
\label{core_radius}
    r_c=\frac{2\pi}{m_a v} \simeq 130\pi\left(\frac{m_a}{10^{-22}~ \mathrm{eV}}\right)^{-1}\left(\frac{v}{10^{-3}}\right)^{-1} \mathrm{pc},
\end{equation}
where $v\sim 10^{-3}$ is the mean velocity of Milky Way DM \cite{Liu:2019brz}. Within the solitonic core $r<r_c$, the ALP field is assumed to be coherent and homogeneous. The simulation shows the DM density profile to be \cite{Schive:2014dra}
\begin{equation}
\label{soliton+NFW}
    \rho_{\mathrm{DM}}=\left\{\begin{array}{ll}190\times \left(\frac{m_a}{10^{-18}~\mathrm{eV}}\right)^{-2}\left(\frac{r_{c}}{1~\mathrm{pc}}\right)^{-4} M_{\odot}~ \mathrm{pc}^{-3}, & \text {for } r<r_{c} \\ \frac{\rho_{0}}{r / R_{g}\left(1+r / R_{g}\right)^{2}}, & \text {for } r>r_{c}\end{array}\right.
\end{equation}
where $\rho_0 = 1.4\times 10^7\,{\rm M}_{\odot}/{\rm kpc}^3$ and $R_g=16$~kpc. Though a smooth transition between the solitionic core and the NFW distribution should exist in reality, we can ignore such a subtlety by considering the emission of the photons from the inner region of the solitonic core.

The linearly polarized source that we consider in this study is around the SMBH, whose size is much smaller than the extent of the solitonic core for $m_a\sim10^{-18}$~eV. One may wonder if the flat profile (\ref{soliton+NFW}) is still valid around the horizon of the SMBH Sgr A$^\star$. Indeed there could be some accumulation effects \cite{Hui:2019aqm} which make the density of the surrounding ALP fields even higher than the one of the solitonic core which is about $10^8$ GeV~${\rm cm}^{-3}$. Thus our adoption of Eq. (\ref{soliton+NFW}) as a benchmark would result in conservative  exclusion limits. There could be some wash-out effects due to the presence of the SMBH, which will be discussed in Section \ref{discussion}.

\section{Data analysis and results}\label{data analysis}
Sgr A$^\star$ is an SMBH with mass of $\sim4.3\times 10^6$~M$_{\odot}$. Its Schwarzschild radius is about $4\times 10^{-7}$~pc (or 0.08 A.U.), given a distance of $\sim8$~kpc away from the Earth. Quite a number of observations revealed the linearly polarized emission from Sgr A$^\star$ \cite{Bower:2004zg,Johnson:2015iwg,Gold:2016hld,Lu:2018uiv}. It has been shown that the position angles vary with time, due possibly to the variation of the magnetic field configuration.

In this work we use the observational data given in Ref.~\cite{Johnson:2015iwg}, which adopted a sub-array of the EHT to map out Sgr A$^\star$ at a resolution of $\sim60 ~\mu as$ (corresponding to a spatial resolution of $\sim 6$ Schwarzschild radii). The observations were carried out for 5 nights. The position angle varied significantly among different days of the observations. Even within one day, there was large variation at different time. These variations reveal that the magnetic field structure is highly dynamic in the region close to the horizon of Sgr A$^\star$. It is thus difficult to use all these data to search for or constrain ALPs, which are expected to give a periodic evolution of the position angle. However, we note that for day 82 of the observations, the position angle remain relatively stable with only small perturbations. Thus we use the data taken on day 82 to search for the ALP-photon coupling effect.

To examine whether there are ALP-photon coupling signals, we fit these datasets with two types of hypotheses, the background hypothesis without ALPs ($H_0$) and the signal hypothesis with ALPs ($H_1$). As we have mentioned before, the background induced effect may be complicated and irregular, making the statistical analysis non-trivial. For the selected day 82 observations, we assume that the background position angle is a constant $\phi_{\textrm{bkg}}$ without significant variations. This assumption is reasonable, otherwise the background should change coincidentally with the ALP induced effect. The ALP induced position angle shift is shown in Eq. (\ref{delta_angle2}). The $\chi^2$ function for both hypotheses is
\begin{equation}\label{chi2}
    \chi^2=\sum_{i=1}^{N} \frac{\left(\phi_{{\rm obs},i}-\phi(t_i)\right)^{2}}{\sigma_{i}^{2}},
\end{equation}
where $\phi_{{\rm obs},i}$ and $\sigma_i$ are the observed position angle and its error. As for the background hypothesis ($H_0$), $\phi(t_i)$ is the background position angle $\phi_\textrm{bkg}$. While for the signal hypothesis ($H_1$), the model-predicted position angle $\phi(t_i)$ is given in Eq.~(\ref{theorial_phi}).

Errors of the data we use result from two major sources: one is systematic uncertainties in the calibration solution estimated to be $\pm 3^{\circ}$\cite{Johnson:2015iwg}; the other is the thermal noise calculated from the real and imaginary parts in the polarization data. For each time, different array configurations with different baselines were applied. Here we combine the three data sets, i.e., the CARMA array, the SMT$_{\rm R}$-CARMA$_{\rm L}$, and the SMT$_{\rm L}$-CARMA$_{\rm R}$ as given in Fig.~S8 of Ref.~\cite{Johnson:2015iwg}, via adding their $\chi^2$ values together. See Appendix for the results for individual data sets, which show small differences from the combined one.

%As for errors of the data, this work considers both the thermal noise and calibration uncertainties, and the uncertainty of calibration, which is the dominant error, is estimated to be $\pm 3^{\circ}$ \cite{Johnson:2015iwg}. \textcolor{red}{incomplete}

\begin{figure}[htbp] %H为当前位置，!htb为忽略美学标准，htbp为浮动图形
\centering 
\includegraphics[width=0.9\textwidth]{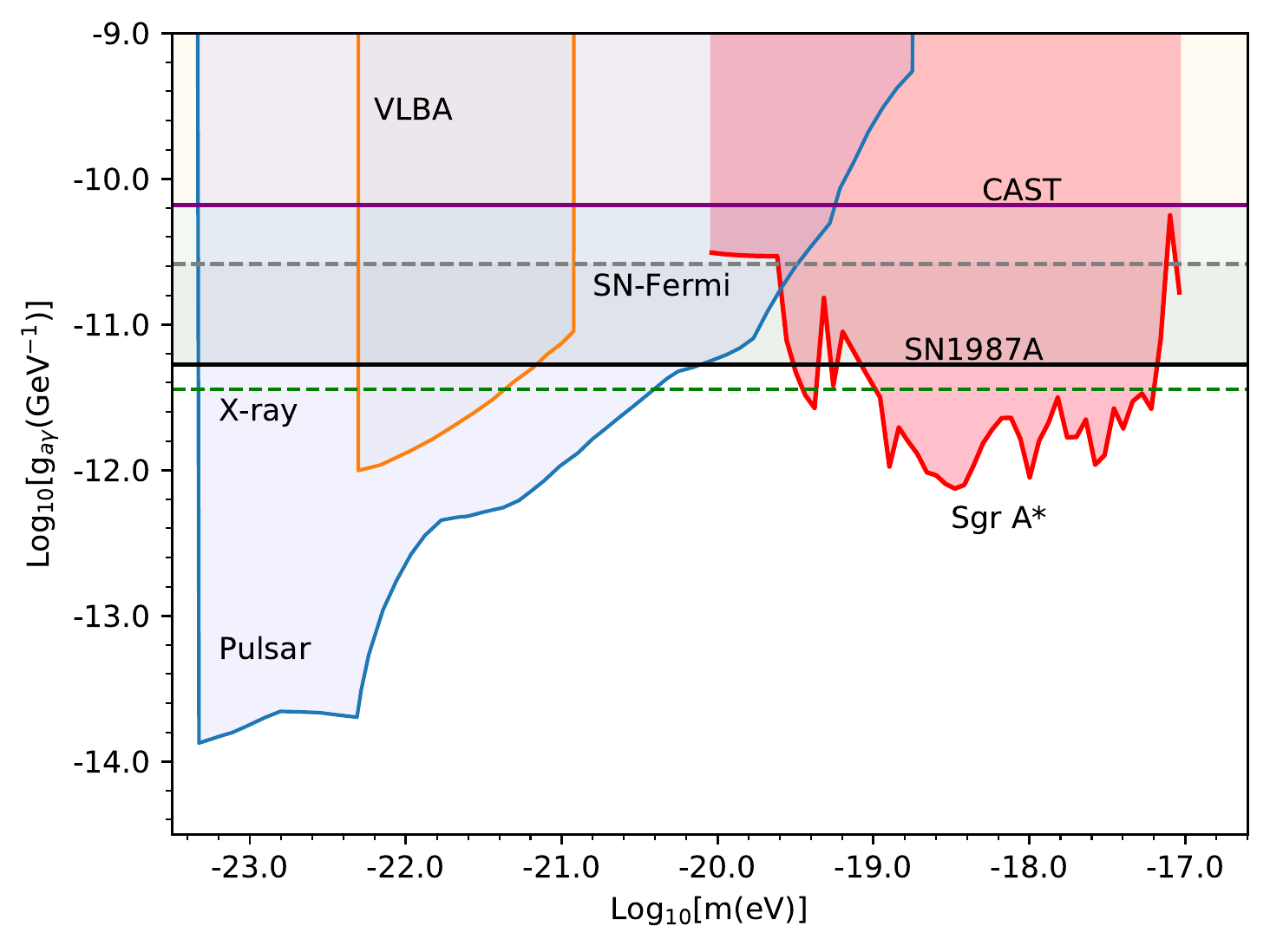} 
\caption{The 95\% confidence level upper limits (red line) of the ALP-photon coupling constant $g_{a\gamma}$ for different ALP mass, obtained using the observations of Sgr A$^\star$ \cite{Johnson:2015iwg}. Other constraints from the CAST experiment \cite{Anastassopoulos:2017ftl} (purple), the supernova SN1987A \cite{Payez:2014xsa} (black), the Fermi-LAT observation of a population of supernovae \cite{Meyer:2020vzy} (grey dashed),  the X-ray observation of super star clusters \cite{Chris:200803305} (green dashed), the VLBA polarization observations of parsec-scale jets from active galaxies \cite{Ivanov:2018byi} (orange) and pulsars \cite{Caputo:2019tms} (blue). %Notice that a cosmological bound also exists in the mass range of this plot based on the assumption that ALP dark matter produced from the misalignment have a lower bound on the decay constant which can translate into $g_{a\gamma}$ \cite{Dror:2020zru}.
} 
\label{Fig.polarization} %用于文内引用的标签
\end{figure}

We define the test statistic (TS) value, TS~$\equiv \chi^2_{H_0} - \chi^2_{H_1}$, to quantify the significance of potential signals from the ALP-photon coupling. The fitting gives a maximum TS value for the combined three data sets of $21.7$, with the best-fit ALP parameters $m_a = 7.9 \times 10^{-18}$ eV and $g_{a \gamma}$= $3.2\times 10^{-11}{\rm GeV^{-1}}$. Such a TS value corresponds to a significance of $\sim4\sigma$ for 3 degrees of freedom ($m_a$, $g_{a\gamma}$, and initial phase $\delta$). Note that the set of best-fit ALP parameters have already been excluded by supernova observations \cite{Payez:2014xsa,Meyer:2020vzy} and the X-ray observation of super star clusters\cite{Chris:200803305}. Therefore, we conclude that no significant ALP signal has been found.
%We calculate TS values of three sets of data, they are 3.04, 2.17 and 0.87 (corresponding to significances of 0.87$\sigma$, 0.61$\sigma$ and 0.21$\sigma$) for TFTG low band, TFTG high band and SGSF low band, respectively. Thus, there is no obvious signal we found.

\begin{figure}[htbp] %H为当前位置，!htb为忽略美学标准，htbp为浮动图形
\centering 
\includegraphics[width=0.9\textwidth]{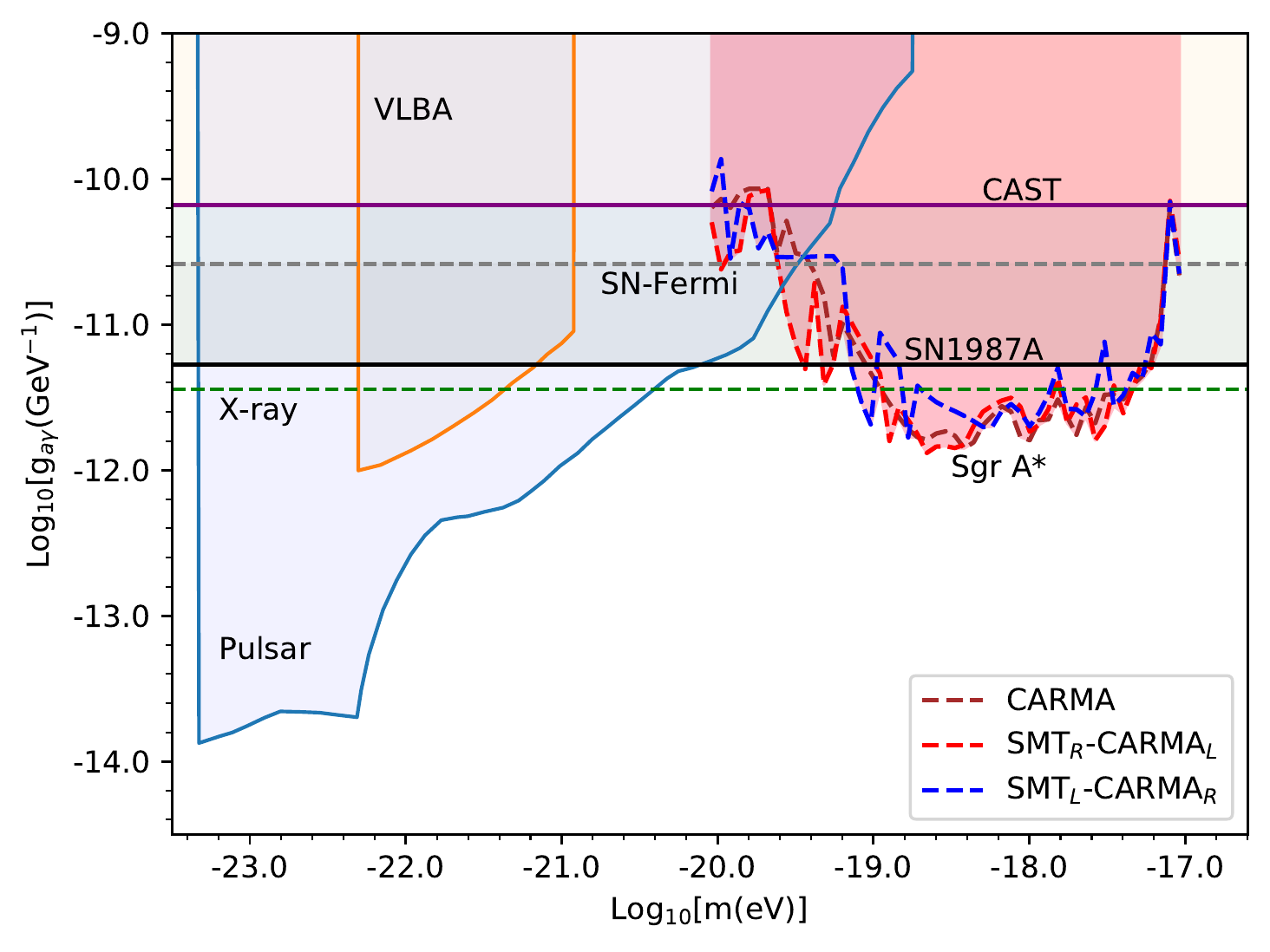}
\caption{Same as Fig.~\ref{Fig.polarization} but for results obtained using the three data sets of Ref.~\cite{Johnson:2015iwg} individually.} 
\label{Fig2.polarization} %用于文内引用的标签
\end{figure}

We then derive the exclusion limits of ALP parameters. 
For a set of fixed ALP mass $m_a$ from $10^{-20}$~eV to $10^{-17}$~eV, we fit the data with different ALP-photon coupling $g_{a \gamma}$ with a scan of $\delta$ in the range of $[0,\,2\pi]$ and $\phi_{\textrm{bkg}}$ in the range of $[0^\circ,\,180^\circ]$. Then we obtain the 95\% confidence level upper limit of the coupling at which the $\chi^2$($g_{a \gamma}$) value is larger by 2.7 than the minimum value. The results are shown in Fig.~\ref{Fig.polarization} and Fig.~\ref{Fig2.polarization}.

For the ALP mass from $\sim10^{-19}$~eV to $6\times10^{-18}$~eV, the constraints from the Sgr A$^\star$ polarization are better than that from the CAST experiment \cite{Anastassopoulos:2017ftl} and supernova \cite{Payez:2014xsa}. The lower limit of the sensitive mass depends on the total observational time, which is about 4 hours, corresponding to $m_a\sim 3\times10^{-19}$~eV according to Eq.~(\ref{oscillation_period}). The upper limit is determined by the observational interval which is about 20 minutes (corresponding to $m_a\sim3\times10^{-18}$~eV). The constraints become weaker for masses far away from the above sensitive region, as it can be seen in Fig.~\ref{Fig.polarization}. Due to the different observational time coverage, the constraints obtained in this work are complementary to those from pulsars as given in Ref.~\cite{Caputo:2019tms}.

We notice that there are some discontinuous jumps in our constraint lines.
For the low ALP mass $m_a$ range from $10^{-20}$ eV to $10^{-18.8}$ eV, these jumps are a result of the degeneracy between the coupling $g_{a\gamma}$ and the background parameter $\phi_\textrm{bkg}$. 
The period $T$ of the ALP induced position angle shift $\Delta\phi (t)$ (signal) is inversely proportional to the ALP mass $m_a$ as shown in Eq.~\ref{oscillation_period}. 
When the ALP mass $m_a$ is smaller $10^{-18.8}$ eV, the half period of the signal $\Delta\phi (t)$ is longer than the duration of observation ($\sim$ 4 hours), and the signal $\Delta\phi (t)$  (proportional to the coupling $g_{a\gamma}$ in Eq.~\ref{delta_angle2}) becomes flat within the observation time, which makes it hard to be distinguished from the background $\phi_\textrm{bkg}$. 
Hence, the signal $\Delta\phi (t)$ in the low mass region could be overestimated with varying degrees, which leads to some discontinuous jumps in the upper limits of the coupling $g_{a\gamma}$. 
While for the high ALP mass region from $10^{-18.5}$ eV to $10^{-17}$ eV, there are some upward jumps induced by some non-significant favored parameter space, of which the most significant one corresponds to the best-fit ALP parameters we obtained ($m_a = 7.9 \times 10^{-18}$ eV and $g_{a \gamma}$= $3.2\times 10^{-11}{\rm GeV^{-1}}$).

\section{Discussion}\label{discussion}
We discuss several potential uncertainties that may affect the analysis in this work. The SMBH in the galactic center may affect the density distribution and the relative phase of the ALP field within the solitonic core. For example, a Kerr black hole with a large spin can result in superradiance of the bosonic field when its Compton wavelength is comparable to the gravitational radius of the black hole \cite{Penrose:1971uk,Press:1972zz,Damour:1976kh,Zouros:1979iw,Detweiler:1980uk,Strafuss:2004qc,Dolan:2007mj,Rosa:2009ei,Dolan:2012yt,Brito:2015oca}). 
A bound state of the bosonic cloud can be built up until the spin of the black hole significantly decreases or the self-interaction of the ALP field becomes important \cite{Yoshino:2012kn,Yoshino:2013ofa,Yoshino:2015nsa}.
The accumulation of the ultra-light bosons could also happen outside a Schwarzschild black hole \cite{Hui:2019aqm}. In both cases, one could expect an enhancement of the ALP density, leading to more stringent constraints of our results.

On the other hand, the presence of the black hole can also induce some washout effect to the birefringence. ALP field may carry angular momentum from superradiance, or if there are vorticities forming \cite{Hui:2020hbq}.
The angular momentum appears in the wave-function as a position-dependent phase $e^{i n \psi}$ where $n$ and $\psi$ refer to azimuthal number and azimuthal angle which is orthogonal to the angular momentum respectively. A spatially dependent phase of the ALP field leads to an average on the shift of the position angle, which may wash out somehow the ALP-induced position angle variation. Without losing generality, taking the azimuthal angle $\psi=0$ and considering the average effect within $\delta\psi$, one obtains the wash-out factor caused by the angular momentum as \cite{Chen:2019fsq}
\begin{equation}
    \frac{1}{\delta \psi} \int_{-\delta \psi / 2}^{\delta \psi / 2} \cos (m_a t+n \psi) d \psi=\frac{\sin (n \delta \psi / 2)}{n \delta \psi / 2} \cos m_a t.
\end{equation}
In reality, the emission region around the SMBH should be asymmetric, and therefore the above wash-out factor should not be exactly zero. Taking the simulation results adopted in Ref.~\cite{Johnson:2015iwg} as an estimate, the emission region occupies about $\delta\psi \sim \pi/2$ of the whole disk, which gives a wash-out factor of only 0.9.
ALPs may also carry a radial dependent phase near the horizon of the black hole \cite{Hui:2019aqm}. Since the Compton wavelength of the ALPs in the mass range relevant in this work is much larger than the size of the black hole horizon, the wash-out effect due to the radial oscillation is negligible. Another effect is that photons may get trapped in the very close vicinity of the black hole, leading to revolving around the black hole. As illustrated by the recent simulation \cite{Johnson:2019ljv}, most of photons travel zero or once around the black hole, before leaving the emission region. Therefore this revolving-induced washout effect should also be small.

Finally, there could be some wash-out along the line of sight direction. The observed radiation comes mainly from a region smaller than the resolution angle of the observations which is a few times of the Schwarzschild radius. When the axion mass is less than $10^{-17}$ eV, its Compton wavelength is much longer than the source emitting size, and the average along the line of sight becomes again negligible.

\section{Summary}\label{summary}
Due to the balance between the gravitational interaction and the quantum pressure, a dense region of ALP field can form in the center of the Milky Way. For linearly polarized photons emitted in the galactic center, the position angles are predicted to oscillate periodically due to the interaction with coherently oscillating ALP field. A polarimetric measurement with good spatial resolution is particularly crucial for such a test. 
%As the development of VLBI technique, especially the EHT project, we will have more chances to precisely observe the SMBHs (such as Sgr A$^\star$, M87$^\star$). 
In this work we use the polarimetric observations of the position angles of millimeter photons from Sgr A$^\star$ taken by a sub-array of EHT \cite{Johnson:2015iwg} to search for the ALP-photon coupling signature. No clear periodic oscillation of the position angles has been found. Using a dark matter density profile with a solitonic core forming inside, we obtain constraints on the ALP-photon coupling constant $g_{a\gamma}$ for ALP mass from $10^{-20}$ eV to $10^{-17}$ eV range (Fig.~\ref{Fig.polarization}). Our results give the strongest constrains on the coupling constant for the mass range $m_a\sim (10^{-19}-10^{-18})$~eV. The upcoming observations of Sgr A$^\star$ by the full EHT array are expected to further improve the sensitivities in searching for ALPs.

\acknowledgments
We thank Michael D. Johnson for providing us with the polarization data of Sgr A$^\star$, and Shuaibo Bian, Yi-Zhong Fan, Lei Feng, Zhaoqiang Shen, Xiao Xue, and Yue Zhao for valuable comments and discussions. Z.Q.X is supported by the National Natural Science Foundation of China (NSFC) under Grants No. 12003069 and No. U1738210.
Y.F.C. is supported in part by NSFC under Grants No. 11722327, No. 11653002, No. 11961131007, No. 11421303, by the CAST-YESS (2016QNRC001), by the National Youth Talents Program of China, and by the Fundamental Research Funds for Central Universities. Y.C. is supported by the China Postdoctoral Science Foundation under Grant No. 2020T130661, No. 2020M680688 and the International Postdoctoral Exchange Fellowship Program. J.S. is supported by the NSFC under Grants No. 11947302, No. 11690022, No. 11851302, No. 11675243 and No. 11761141011, and by the Strategic Priority Research Program of the Chinese Academy of Sciences under Grants No. XDB21010200 and No. XDB23000000. Q.Y. is supported by the NSFC under Grants No. 11722328, No. 11851305, the 100 Talents program of Chinese Academy of Sciences, and the Program for Innovative Talents and Entrepreneur in Jiangsu.

% \bibliographystyle{plain}
%\bibliographystyle{unsrt}
% \bibliography{references}

\bibliographystyle{JHEP}

\section*{Appendix: Individual constraints from three data sets}
Here, we provide constraints on $g_{a\gamma}$ using the three data sets individually, i.e., the CARMA array, the SMT$_{\rm R}$-CARMA$_{\rm L}$, and the SMT$_{\rm L}$-CARMA$_{\rm R}$ as given in Fig.~S8 of Ref.~\cite{Johnson:2015iwg}. Compared with the combined results of Fig.~\ref{Fig.polarization}, the individual constrains are slightly weaker. 

\end{document}